\documentclass[aps,prd,10pt]{revtex4-2}
\usepackage{amsmath}
\usepackage{amssymb}
\usepackage{amsthm}
\usepackage{graphicx}
\usepackage{braket}
\usepackage{multirow}
\usepackage{hyperref}
\usepackage{xcolor}

\begin{document}

\title{Inflationary Gravitational Waves and Laboratory Searches\\ as Complementary Probes of Right-handed Neutrinos}

\author{Zafri A. Borboruah}
\email{zafri123@iitb.ac.in}
\affiliation{Indian Institute of Technology Bombay, Mumbai 400076, Maharashtra, India}

\author{Frank F. Deppisch}
\email{f.deppisch@ucl.ac.uk;}
\affiliation{Department of Physics and Astronomy, University College London, Gower Street, London, WC1E 6BT, United Kingdom}

\author{Anish Ghoshal}
\email{anish.ghoshal@fuw.edu.pl}
\affiliation{Institute of Theoretical Physics, Faculty of Physics, University of Warsaw, ul. Pasteura 5, 02-093 Warsaw, Poland}

\author{Lekhika Malhotra}
\email{lekhika.malhotra@iitb.ac.in}
\affiliation{Indian Institute of Technology Bombay, Mumbai 400076, Maharashtra, India}

\begin{abstract}
We analyze the damping of inflationary gravitational waves (GW) that re-enter the Hubble horizon before or during a post-inflationary era dominated by a meta-stable, right-handed neutrino (RHN), whose out-of-equilibrium decay releases entropy. Within a minimal type-I seesaw extension of the Standard Model (SM), we explore the conditions under which the population of thermally produced RHNs remain long-lived and cause a period of matter-domination. We find that the suppression of the GW spectrum occurs above a characteristic frequency determined by the RHN mass and active-sterile mixing. For RHN masses in the range $0.1$–$10$~GeV and mixing $10^{-12} \lesssim |V_{eN}|^2 \lesssim 10^{-5}$, we estimate such characteristic frequencies and the signal-to-noise ratio to assess the detection prospects in GW observatories such as  THEIA, $\mu$-ARES, LISA, BBO and ET. Additionally we use LIGO data to put upper bounds on the reheating temperature after inflation, for a given blue-tilted GW spectrum. We find complementarity between GW signals and laboratory searches in SHiP, DUNE and LEGEND-1000. Notably, RHN masses of $0.2$–$2$~GeV and mixing $10^{-10} \lesssim |V_{eN}|^2 \lesssim 10^{-7}$ are testable in both laboratory experiments and GW observations. Additionally, GW experiments can probe the canonical seesaw regime of light neutrino mass generation, a region largely inaccessible to laboratory searches. 
\end{abstract}

\maketitle

\section{Introduction}
All fermionic particles in the Standard Model (SM) of particle physics come in left- and right-handed variants except for the neutrinos which are only left-handed. This may be related to the fact that we observe neutrinos only through their left-handed SM interactions unlike the other particles. However, the SM fails to explain the masses of the neutrinos as evidenced by neutrino oscillation experiments~\cite{Davis:1994jw, Super-Kamiokande:1998kpq, KamLAND:2002uet, SNO:2002tuh}. The tiny neutrino masses $m_\nu \lesssim 0.1$~eV can not be generated through the traditional Higgs mechanism since the Yukawa couplings between the left and right-handed neutrinos required for generating sub-eV Dirac masses turn out to be of the order $10^{-12}$, which is not natural. Therefore, we might look into the possibility that neutrinos are of Majorana nature. The simplest way of Majorana neutrino mass generation is through the well-known type-I seesaw mechanism \cite{Minkowski:1977sc, Mohapatra:1979ia, Gell-Mann:1979vob, Yanagida:1979as, Glashow:1979nm, Schechter:1980gr, 10.1143/PTP.64.1103, PhysRevLett.44.912}, with heavy RHNs. If the RHNs are sterile, i.e., uncharged under any gauge interactions, then the only coupling associated with them is through the Yukawa interaction with a left-handed lepton doublet and the Higgs doublet. If the heavy RHNs required by a viable type-I seesaw framework have masses in the MeV to TeV scale, they can be searched for in laboratory experiments, such as beta decays, meson decays, beam dump experiments and at colliders. For a comprehensive summary of such experiments, along with the existing constraints and future prospects of RHN searches, see, e.g., \cite{Atre:2009rg, Bolton:2019pcu, Abdullahi:2022jlv}. These RHN states are vastly heavier than the light active neutrinos, hence they are also referred to as heavy neutral leptons (HNLs).

The detection of gravitational waves (GWs) from black hole mergers by the LIGO and Virgo collaboration \cite{LIGOScientific:2016aoc, LIGOScientific:2016sjg} and strong evidence for the presence of a stochastic GW background (SGWB) from measurements carried out by several pulsar timing array (PTA) collaborations \cite{Carilli:2004nx, Janssen:2014dka, Weltman:2018zrl, EPTA:2015qep, EPTA:2015gke, NANOGrav:2023gor, NANOGrav:2023hvm} have lead to postulations of BSM new physics scenarios associated with GW production in the early Universe, including primordial origins of GWs related to cosmic inflation. Our work here proposes to form a synergy between GW observations and laboratory experiments involving seesaw and HNL searches. We will investigate quantum tensor fluctuations of the metric during cosmic inflation which propagate as Gravitational Waves in the post-inflationary era. These GW modes go out of Hubble horizon during inflation~\cite{Grishchuk:1974ny, Starobinsky:1979ty, Rubakov:1982df, Guzzetti:2016mkm}. Upon their horizon re-entry after inflation, these primordial gravitational waves (PGWs) can act as a book-keeping element of the cosmic expansion history of our Universe~\cite{Seto:2003kc, Boyle:2005se, Boyle:2007zx, Kuroyanagi:2008ye, Nakayama:2009ce, Kuroyanagi:2013ns, Jinno:2013xqa, Saikawa:2018rcs, Chen:2024roo}, since they propagate across all cosmological epochs and reach us today. Any deviation from the standard radiation-dominated pre-BBN history can be inferred by studying the features in the PGW spectral shapes since the expansion history of the Universe determines the red-shifting of the GW amplitudes and frequencies to the present day~\cite{Nakayama:2008ip, Nakayama:2008wy, Kuroyanagi:2011fy, Buchmuller:2013lra, Buchmuller:2013dja, Jinno:2014qka, Kuroyanagi:2014qza, Bernal:2020ywq, Ghoshal:2023sfa, Ghoshal:2022ruy, Chen:2024roo}. In this regard, we consider an epoch where HNLs dominate the energy budget of the Universe, thus affecting its post-inflationary evolution which leads to characteristic GW spectral shapes testable in various current and upcoming GW detectors. Such primordial features in the GW spectrum usually reveals two important quantities: (a) the time when HNL matter domination began which is set by the highest frequency that departs from a flat spectrum, and (b) the duration of HNL matter domination which corresponds to the width of the feature in frequency space~\cite{Berbig:2023yyy, Borboruah:2024eha, Borboruah:2024eal, Ghoshal:2024gai, Bernal:2020ywq, Datta:2022tab, Datta:2023vbs, Chianese:2024nyw, Ghoshal:2022ruy, Datta:2025yow}.  

Long-lived HNLs are notoriously difficult to detect in conventional collider experiments due to their displaced vertices, which render them effectively invisible~\cite{Feng:2022inv}. For HNLs with masses in the range $0.1~\text{GeV} \lesssim 10$~GeV, beam dump, long-lived particle search experiments, such as Deep Underground Neutrino Experiment (DUNE)~\cite{DUNE:2015lol, Berryman:2019dme}, MAssive Timing Hodoscope for Ultra-Stable neutraL pArticles (MATHUSLA)~\cite{Curtin:2018mvb, Curtin:2023skh} and Search for Hidden Particles (SHiP)~\cite{SHiP:2015vad} offer promising prospects for probing large regions of the parameter space. Nevertheless, even if an HNL is detected, the information extracted is typically limited to its mass and lifetime, leaving the rest of the hidden sector poorly constrained. It is therefore of significant importance to find complementary sources of information about such long-lived HNLs. We propose, in this paper, that primordial features in the SGWB from the inflationary paradigm will be useful in obtaining independent evidence for and information on HNLs. This suggests a potential synergy between GW observatories and laboratory experiments in testing such a beyond-the-Standard-Model (BSM) scenario.

The paper is organized as follows: In Sec.~\ref{sec:HNL searches}, we give a review of type-I seesaw and HNL decays along with current and future HNL searches. In Sec.~\ref{sec:HNL domination}, we discuss the conditions for HNL domination in the early Universe and Sec.~\ref{sec:IGW} covers the suppression of primordial gravitational waves of inflationary origin due to HNL domination and we show the future projections of deviations in signal-to-noise ratio of the suppressed GW spectra due to HNL domination as compared to the standard spectra for various GW experiments. We demonstrate the complementarity of GW searches with laboratory searches in the HNL parameter space. Finally, in Sec.~\ref{sec:conclusion}, we conclude and discuss future implications. 


\section{Heavy neutral leptons}
\label{sec:HNL searches}

As the underlying model, we consider the conventional type-I seesaw with three HNLs $N_i$,
\begin{align}
\label{eq:seesaw}
    \mathcal{L} = \sum_{\alpha,i} y_{\alpha i} \overline{L_\alpha} (i\sigma_2) H^\dagger N_i + \frac{1}{2} \sum_i m_{N_i} \overline{N^c_i} N_i + \text{H.c.}.
\end{align}
Here, $L_\alpha$ are the SM lepton doublets ($\alpha = e, \mu, \tau$), $H$ is the SM Higgs doublet and $\sigma_2$ is the second Pauli matrix. We assume, without loss of generality that the symmetric RHN mass matrix is diagonal, $M_N = \text{diag}(m_{N_1}, m_{N_2}, m_{N_3})$. After electroweak symmetry breaking with $\braket{H}\equiv v \sim 174$~GeV, the light active neutrinos acquire and effective Majorana mass matrix $M_\nu = - m_D \cdot M_N^{-1} \cdot m_D^T$, at leading order in the seesaw expansion with $m_D \equiv y v \ll M_N$. It is diagonalized by the PMNS mixing matrix $U_\text{PMNS}$, constrained by measurements of neutrino oscillations. The RHN Majorana mass term is a priori unconstrained, as it is unrelated to SM electroweak symmetry breaking; it can in principle have any scale from eV and below to the Planck scale. We here consider HNLs with masses in the range $0.1~\text{GeV} \leq m_N \leq 10$~GeV as the regime of interest for direct searches and motivated by viable low-scale leptogenesis.

In the following, we work in a simplified scenario with a single HNL denoted by $N$. We ignore any flavor structure and we assume that the HNL dominantly couples to the electron-flavor active neutrino and the Lagrangian \eqref{eq:seesaw} simplifies to 
\begin{align}
\label{eq:seesawsimple}
    \mathcal{L} = y_{e N} \overline{L_e} (i\sigma_2) H^\dagger N 
    + \frac{1}{2} m_N \overline{N^c} N + \text{h.c.}.
\end{align}
A light active neutrino mass scale $m_\nu = |V_{e N}|^2 m_N$ is then generated, where $V_{e N}$ is the admixture between the active neutrino and the HNL, $V_{e N} = y_{e N} v / m_N$. We treat $m_N$ and $|V_{e N}|^2$ as free parameters. The focus of our paper is to study the interplay between the constraints and future sensitivity reaches from laboratory HNL searches and GW detectors. We do not attempt to consider the full light neutrino mass spectrum and the PMNS mixing pattern observed in oscillations. Instead, we take the limit on the effective $\beta$ decay mass, $m_\beta < 0.45$~eV at 90~\% confidence level~(CL), from KATRIN \cite{Katrin:2024tvg} as an upper limit on $m_\nu$, whereas solar neutrino oscillations point toward a smallest scale $m_\nu > \sqrt{\Delta m^2_\text{sol}} \approx 9\times 10^{-3}$~eV~\cite{Esteban:2020cvm}. Thus, successful neutrino mass generation favours a parameter space of interest $9\times 10^{-3}~\text{eV} < |V_{e N}|^2 m_N < 0.45$.

For GeV-scale HNLs, this requires very small active-sterile mixing strengths,
\begin{align}
\label{eq:mixing}
    |V_{e N}|^2 = 
    \frac{m_\nu}{m_N} \lesssim 10^{-10} \left(\frac{1~\text{GeV}}{m_N}\right),
\end{align}
resulting in long-lived HNLs that can be searched through their displaced vertices in laboratory experiments. On the other hand, if such long-lived HNLs are present in the early Universe, they can trigger a period of intermediate matter domination that can imprint on the GW spectrum arising due to tensor perturbations generated during cosmic inflation. We should emphasize, though, that the relation in Eq.~\eqref{eq:mixing} is not strictly required. Smaller mixing is possible if there are other contributions to the light neutrino masses, such as from additional HNLs. Likewise, the mixing can be larger if lepton number is approximately conserved. This is not only achieved in the limits $m_N\to 0$ and $m_N\to\infty$ but also if pairs of HNLs form Dirac fermion states themselves. By feebly violating lepton number symmetry, e.g., through a Majorana mass term $\mu \ll m_N$, light neutrino masses are generated, $|m_\nu| \sim (y_\nu v/m_N)^2 \mu = |V_{\ell N}|^2 \mu$, while the HNLs form quasi-Dirac pairs with a small mass splitting $\Delta m_N \sim \mu$.

\subsection{HNL decays and direct searches}
\label{sec:HNL decay}

This section summarizes the most relevant HNL decay channels, considering both charged and neutral current interactions as mediators. Most results for sufficiently light HNLs are already established in the literature \cite{Atre:2009rg, Bolton:2019pcu, Abdullahi:2022jlv} and we use the partial decay widths summarized in \cite{Feng:2024zfe}. Two basic processes  contribute to the decay modes: charged current decays leading to final particles such as lepton pairs ($\nu_e, e$) or up/down quark pairs ($u, d$), and neutral current decays, leading to fermion pairs $f\bar f$. For the decays $N \to \nu_e e^- e^+$ and $N \to \nu_e \nu_e \bar{\nu}_e$, both processes can contribute, with interference. The general formula for charged current-mediated processes, such as $N \to e^- \nu_\beta \ell_\beta^+$ ($\beta \neq e$) and $N\to e u_i \bar{d}_j$, is
\begin{align}
    \Gamma(N \to e^- u \bar{d}) = \frac{G_F^2 m_N^5}{192\pi^3} |V_{e N}|^2 
    I(x_u, x_d, x_e),
\end{align}
where $G_F = 1.166\times10^{-5}$ GeV$^{-2}$ is the Fermi constant and $x_e = \frac{m_e}{m_N}$, $x_u = \frac{m_u}{m_N}$, $x_d = \frac{m_d}{m_N}$. The function $I(x_u, x_d, x_l)$ accounts for corrections due to finite masses of the final-state fermions. The full expression is given in \cite{Atre:2009rg, Barman:2022scg}. Likewise, the decay width for neutral current-mediated decays $N \to \nu_e f \bar{f}$ depends on the types of final fermions. For the case of pure neutrino final states, the decay width simplifies to
\begin{align}
    \Gamma(N \to \nu_e \nu_\beta \bar\nu_\beta) 
    = (1 + \delta_{e\beta}) \frac{G_F^2 m_N^5}{768\pi^3} |V_{eN}|^2.
\end{align}

At masses below the QCD scale ($m_N \lesssim \Lambda_{\rm QCD}$), the decay products form single mesons, and decay modes involving charged and neutral mesons are considered. The decay widths for these processes are derived for charged pseudoscalar mesons ($\pi^\pm, K^\pm$), neutral pseudoscalar mesons ($\pi^0, \eta$), charged vector mesons ($\rho^\pm$, $a_1^\pm$) and neutral vector mesons ($\rho^0$, $a_1^0$). For HNLs with masses above 1~GeV, multi-hadron final states become kinematically accessible. The total hadronic decay width, $\Gamma_{\rm had}$ is estimated by comparing the decay width into quarks and the corresponding loop corrections, which are similar to those found for $\tau$-lepton decays. QCD corrections are expected to be less than $30\%$ for HNLs with $m_N \gtrsim 1$~GeV. Overall, the full hadronic decay width dominates for $m_N \gtrsim 1$ GeV.

Corresponding to their total decay width, HNLs are naturally long-lived if light compared to the mediating SM gauge bosons and for the small active-sterile mixing strength expected for successful light neutrino mass generation,
\begin{align}
    L_N \approx 25~\text{mm}
    \cdot\frac{10^{-10}}{|V_{\ell N}|^2}
    \cdot\left(\frac{10~\text{GeV}}{m_N}\right)^2.
\end{align}
This approximation is roughly valid for $1~\text{GeV} \lesssim m_N \lesssim m_W$. HNLs lighter than a few GeV can be abundantly produced in beam dump experiments through meson decays, resulting in displaced vertex signatures that can be searched for with small background rates. As representative examples, we will highlight the sensitivity of DUNE~\cite{DUNE:2015lol, Berryman:2019dme}, MATHUSLA~\cite{Curtin:2018mvb, Curtin:2023skh} and SHiP~\cite{SHiP:2015vad}, to compare with GW observations in the framework considered here. The strong sensitivity of such high-luminosity, displaced-vertex searches will be shown in Figs.~\ref{fig:specBP}~(right) and \ref{fig:SNR1}, which shows the future sensitivities as well as current constraints on the active-sterile mixing strength $|V_{eN}|^2$ as function of the HNL mass $m_N$.

\subsection{Neutrinoless double beta decay}
\label{sec:0vbb}

The most sensitive probe of the Majorana nature of light active neutrinos is $0\nu\beta\beta$ decay~\cite{Agostini:2022zub}. In addition, $0\nu\beta\beta$ decay is sensitive to other exotic sources of lepton number violation \cite{Doi:1981, Cirigliano:2017djv, Graf:2018ozy, Deppisch:2020ztt, Deppisch:2020oyx, Bolton:2020ncv, Bolton:2021pey, Agostini:2022bjh}. This includes the exchange of Majorana HNLs. In the simplified scenario of a single HNL mixing with electron flavor, the decay half-life $T_{1/2}^{0\nu}$ for $m_N \gtrsim 100$~MeV is approximately \cite{Bolton:2022pyf}
\begin{align}
\label{eq:0vbb:half-life-heavy}
    \frac{10^{28}~\text{yr}}{T_{1/2}^{0\nu}} \approx 
    \left(\frac{|V_{eN}|^2}{10^{-9}}\cdot\frac{1~\text{GeV}}{m_N}\right)^2.
\end{align}
Here, nuclear matrix elements for the isotope $^{76}$Ge~\cite{Deppisch:2020ztt} are used. The dependence is modified around the $0\nu\beta\beta$ momentum scale $\approx 100$~MeV and at the crossover, the momentum dependence should be accounted for more carefully~\cite{Babic:2018ikc, Dekens:2020ttz}. The sensitivity in Eq.~\eqref{eq:0vbb:half-life-heavy} is compared with $T^{0\nu}_{1/2}(^{76}\text{Ge}) = 10^{28}$~yr, the projected reach of Large Enriched Germanium Experiment for Neutrinoless double-beta Decay (LEGEND)-1000~\cite{Zsigmond:2020bfx}. Current experimental limits are of the order $T^{0\nu}_{1/2} \gtrsim 10^{26}$~yr~\cite{PhysRevLett.117.082503, GERDA:2020xhi}. In Figs.~\ref{fig:specBP}~(right) and \ref{fig:SNR1}, we likewise show the sensitivity of $0\nu\beta\beta$ decay searches to HNLs, in comparison with direct searches and GW observations.


\section{HNL domination in the early Universe}
\label{sec:HNL domination}

As discussed above, the HNL can interact with the SM plasma due to the active-sterile mixing via scattering, annihilation and decays. At temperatures much higher than $m_N$, the interaction rate is approximated as $\Gamma_N^{\rm int}\sim G_F^2T^5|V_{eN}|^2$. The HNL will be thermally produced if the following condition is satisfied~\cite{Boyarsky:2020dzc, Sabti:2020yrt, Syvolap:2021yan, Notzold:1987ik, Dolgov:2000ew},
\begin{equation}
\label{eq:nonThCond1}
    \Gamma_N^{\rm int}(T_{\rm max})>3\mathcal{H}(T_{\rm max}),
\end{equation}
where $\mathcal{H}(T)=\sqrt{\frac{8\pi^3g_\star(T)}{90}}\frac{T^2}{M_{\rm Pl}}$ is the Hubble rate during radiation domination with $g_*$ being the relativistic degrees of freedom and $M_{\rm Pl}=1.22\times10^{19}$ GeV being the Planck mass. The factor 3 is usually taken in cosmology because in an expanding Universe, the interaction rates should be significantly larger than the Hubble rate in order to bring the interacting species to thermal equilibrium. $T_{\rm max}$ is the temperature when $N$ production is maximum, given by~\cite{Syvolap:2021yan,Boyarsky:2009ix,Dodelson:1993je}
\begin{equation}
    T_{\rm max}\sim10\,\text{GeV}\left(\frac{m_N}{1 \text{\,GeV}}\right)^{1/3}.
\end{equation}
This is derived by considering temperature and momentum-dependent mixing strength of the HNL below electroweak scale~\cite{Barbieri:1989ti,Notzold:1987ik}. In this paper, we are only interested in thermally produced HNLs. This gives a lower bound on the mixing strength above which $N$ can enter thermal equilibrium~\cite{Syvolap:2021yan},
\begin{equation}
\label{eq:nonThCond}
    |V_{eN}|^2\gtrsim 6\times10^{-12}\left(\frac{1\,\text{GeV}}{m_N}\right).
\end{equation}
\begin{figure}[t!]
    \centering
    \includegraphics[width=0.6\linewidth]{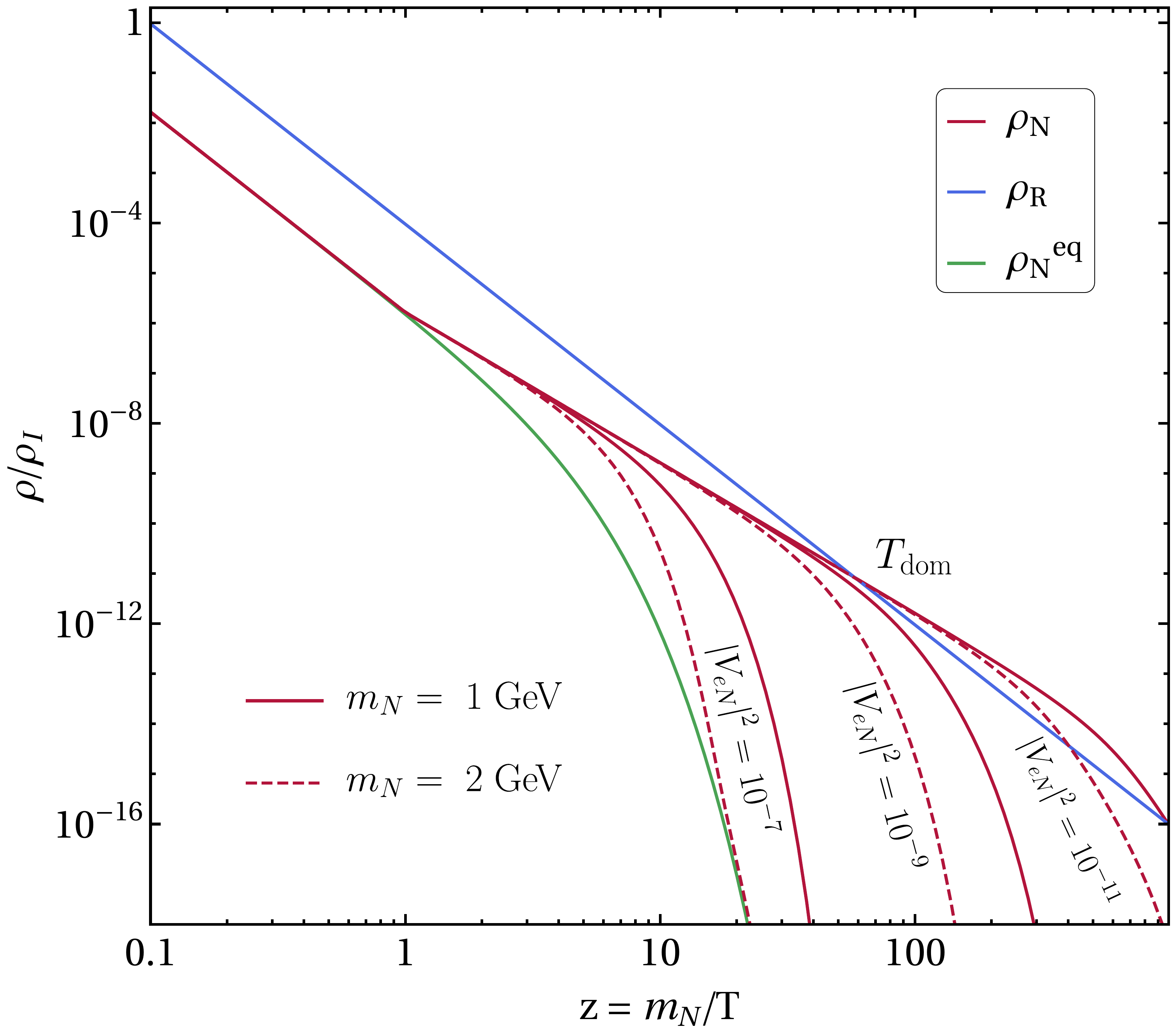}
    \caption{HNL energy density evolution, obtained by solving the Boltzmann Eqs.~\eqref{eq:boltzrho}, for varying $m_N$ and $|V_{eN}|^2$ as depicted (red solid and dashed curves). For comparison, the green curve represents the HNL equilibrium energy density whereas the blue curve shows the radiation energy density. $T_\text{dom}$ indicates the temperature of intermediate HNL matter-domination with mass $m_N=1$~GeV as an example.}
    \label{fig:densityBP}
\end{figure}
After production, $N$ remains in equilibrium until its interaction with SM plasma freezes out at temperature $T_f$. The evolution of the energy densities of $N$ and radiation at temperatures below $T_f$ are estimated by the following set of coupled Boltzmann equations,
\begin{align}
\label{eq:boltzrho}
    \dot{\rho}_{N} &= 
    - 3\mathcal{H}\left(\rho_{N}+p_{N}\right) 
    - \braket{\Gamma_{N}}(\rho_{N}-\rho_{N}^{eq}), \nonumber\\
    \dot\rho_{R} &= 
    - 4\mathcal{H}\rho_{R} 
    + \braket{\Gamma_{N}}(\rho_{N}-\rho_{N}^{eq}),
\end{align}
where $\braket{\Gamma_N}$ is the thermally averaged total decay width of $N$. We ignored the scattering and annihilation terms because they are frozen out below $T_f$. The dot denotes differentiation with respect to cosmic time. Here $\rho_N$ and $\rho_N^{\text{eq}}$ denote the energy density and equilibrium energy density of $N$, respectively, while $\rho_R$ corresponds to the radiation energy density of the Universe. The term $p_N$ represents the pressure of the HNL gas. To account for the transition between relativistic (radiation-like) and non-relativistic (matter-like) regimes of $N$ evolution, we assume $p_{N} = \rho_{N}/3$ for $T \gtrsim m_N$ and $p_{N} = 0$ for $T \lesssim m_N$, where $T$ is the temperature of the Universe.

In Fig.~\ref{fig:densityBP}, we show the solutions of the Boltzmann equations. The blue solid line illustrates the evolution of the radiation energy density as a function of the parameter $z = m_N/T$. The red lines depict the evolution of the energy density of $N$, while the green solid line represent the equilibrium energy density. The values of $m_N$ and $|V_{eN}|^2$ are given within the plot. For $z < 1$, $N$ behaves as radiation, and its energy density evolves as $a^{-4}$. For $z > 1$, $N$ become non-relativistic and its energy density evolves as matter ($\propto a^{-3}$). We consider a thermal abundance of $N$ at $z=0.1$ (which means Eq.~\eqref{eq:nonThCond} is satisfied) and demonstrate the effect of $m_N$ and $|V_{eN}|^2$ on the evolution.

As shown in the plot, increasing the value of $|V_{eN}|^2$ leads to an earlier decay of $N$. Before decaying, $N$ becomes non-relativistic as long as its total decay width is smaller than the Hubble rate at temperatures around $m_N$, i.e. $\Gamma_N\ll \mathcal{H}(m_N)$. Assuming that the decay is instantaneous and the decay products thermalize with the radiation bath quickly, the decay temperature $T_{\rm dec}$ can be estimated from the condition $\Gamma_N\sim\mathcal{H}(T_{\rm dec})$ as
\begin{equation}
\label{eq:Tdec}
    T_{\rm dec}\sim \left(\frac{90}{8\pi^3g_\ast(T_{\rm dec})}\right)^\frac{1}{4}\sqrt{\Gamma_N\, M_{\rm Pl}}.
\end{equation}

\begin{figure}[t!]
    \centering
    \includegraphics[width=0.9\linewidth]{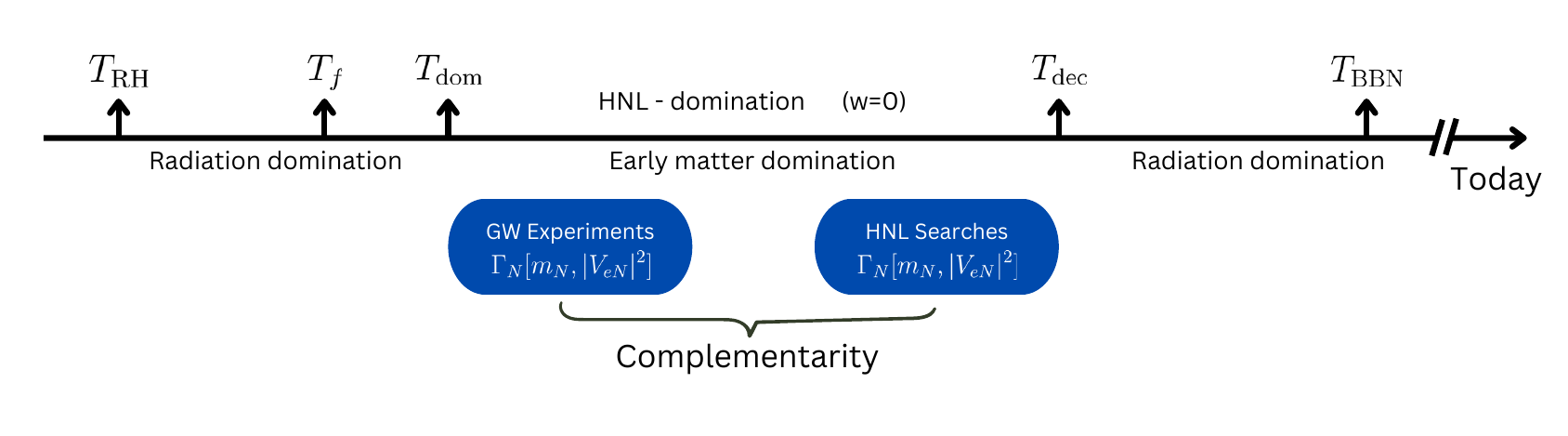}
    \caption{Schematic timeline of key events and eras of cosmic evolution. After reheating ends at $T_\text{RH}$, HNLs are produced thermally from the SM plasma. The interactions decouple and HNLs freeze-out at temperature $T_f$. Eventually, HNLs become non-relativistic at temperatures comparable to the HNL mass, $T \approx m_N$. At $T_\text{dom}$, HNLs start to dominate the energy density of the Universe. This gives rise to a period of matter-domination until the HNLs decay at around $T_\text{dec}$, kick-starting a second radiation dominated epoch before BBN. The modified expansion of the Universe during the HNL-dominated era suppresses the GW spectrum from inflation, depending on $m_N$ and the active-sterile mixing $|V_{eN}|^2$. This effect can be detected in future GW experiments, complementing HNL laboratory searches.}
    \label{fig:schematics}
\end{figure}

For sufficiently small values of $m_N$ and $|V_{eN}|^2$, the energy density of $N$ eventually overtakes the radiation energy density at temperature $T_{\rm dom}$ as shown in the plot for $|V_{eN}|^2=10^{-11}$, giving rise to an early matter-dominated epoch. For an initial thermal abundance of $N$, one can calculate $T_{\rm dom}$ as~\cite{Berbig:2023yyy, Giudice:1999fb}
\begin{equation}
\label{eq:Tdom}
    T_{\rm dom}\sim \frac{7}{4}\frac{m_N}{g_\ast(T_{\rm dom})}\sim 2\%\,m_N.
\end{equation} 
We require $T_{\rm dec}<T_{\rm dom}$ for the existence of the HNL-dominated epoch. The decay of the non-relativistic dominating HNL species into SM particles ends the matter-domination, injecting entropy into the Universe and diluting all energy densities. The effect of this dilution is seen as suppression of inflationary gravitational waves as we will discuss in the next section. Fig.~\ref{fig:schematics} shows the timeline of key events.

The effect on GW spectrum depends on the dilution factor $\mathcal{D}$ which is the ratio of co-moving entropy density after and before the decay event. Assuming the decay products of $N$ thermalize quickly with the radiation bath, the dilution factor at the end of the $N$-dominated epoch can be approximated as~\cite{Scherrer:1984fd, Berbig:2023yyy, Borboruah:2024eha},
\begin{equation}
\label{eq:dilution_factor}
    \mathcal{D} = \frac{s(T_\text{after})a^3(T_\text{after})}{s(T_\text{before})a^3(T_\text{before})}=\left(1+2.95\left(\frac{2\pi^2 \left<g_*(T)\right>}{45}\right)^{\frac{1}{3}}\frac{(\frac{n_N}{s}m_N)^{\frac{4}{3}}}{(M_\text{Pl}\Gamma_N)^{\frac{2}{3}}}\right)^{\frac{3}{4}},
\end{equation}
where $s$ and $a$ are the entropy density and the scale factor respectively. $T_{\rm before/after}$ represents the temperature just before/after the decay. The initial abundance of $N$ is given as the ratio $n_N/s$. We assume that $N$ freezes out while still being relativistic in the parameter range considered. Hence the maximum value of $N$ abundance is~\cite{Bezrukov:2009th},
\begin{equation}
    \label{eq:Nabundance}
    \frac{n_N}{s}=\frac{135\,\zeta(3)}{4\pi^4\,g_{\ast S}},
\end{equation}
where $g_{*S}$ is the number of relativistic degrees of freedom contributing to entropy density and we consider it to be $\sim106.75$ for our analysis. We will use this dilution factor as well as the decay temperature to calculate the GW spectra later.
%


\section{Primordial Gravitational Waves}
\label{sec:IGW}

The early Universe underwent an exponential expansion phase known as inflation. This phase was followed by the reheating process, which generated a high-energy plasma consistent with the Standard Model of particle physics. During inflation, primordial gravitational waves (PGWs) emerged as quantum tensor fluctuations in the metric. Initially, while outside the cosmological horizon, these GWs maintained constant amplitudes. However, once the GWs re-entered the horizon during the radiation-dominated epoch, their amplitudes experienced damping. The propagation of GW modes that re-enter the horizon is characterized by a transfer function. We follow Refs. \cite{Berbig:2023yyy,Borboruah:2024eal,Borboruah:2024eha,Ghoshal:2024gai} to present the standard GW spectrum and then introduce a new concept of relative SNR to disentangle signals of EMD.

\subsection{Gravitational waves spectrum}

The energy density spectrum of the GWs, denoted as $\Omega_{\text{GW}}(k)$, depends on the wave number $k = 2\pi f$, where $f$ is the frequency of the GWs \cite{Berbig:2023yyy}. It is given as
\begin{equation}
\label{eq:OmegaGW}
    \Omega_{\text{GW}}(k) = \frac{1}{12} \left(\frac{k}{a_0 H_0}\right)^2 T_T^2(k) P_T^{\text{prim.}}(k),
\end{equation}
where $a_0 = 1$ is the present-day scale factor, and $H_0 \approx 2.2 \times 10^{-4} \ \text{Mpc}^{-1}$ represents the current Hubble expansion rate \cite{Datta:2022tab}. Here, $P_T^{\text{prim.}}(k)$ is the primordial tensor power spectrum, and $T_T^2(k)$ is the transfer function describing the evolution of GWs as they propagate through a Friedmann-Lemaitre-Robertson-Walker (FLRW) background. The horizon re-entry temperature $T_{\text{in}}$ is expressed as \cite{Nakayama:2008wy},
\begin{equation}
    T_{\text{in}} = 5.8 \times 10^6 \ \text{GeV} \left(\frac{106.75}{g_*(T_{\text{in}})}\right)^{1/6} \left(\frac{k}{10^{14} \ \text{Mpc}^{-1}}\right),
\end{equation}
where $g_*(T_{\text{in}})$ is the effective number of relativistic degrees of freedom at temperature $T_{\text{in}}$.

The primordial tensor power spectrum $P_T^{\text{prim.}}(k)$ can be parametrized in terms of its amplitude $A_T$ and spectral index $n_T$, defined at the pivot scale $k_* = 0.05 \, \text{Mpc}^{-1}$ \cite{Planck:2018jri},
\begin{equation}
    P_T^{\text{prim.}}(k) = A_T(k_*) \left(\frac{k}{k_*}\right)^{n_T}.
\end{equation}
The amplitude $A_T(k_*)$ is related to the scalar power spectrum $A_S(k_*)$ and the tensor-to-scalar ratio $r$, with an upper bound $r \leq 0.035$ from BICEP/Keck observations \cite{BICEP:2021xfz},
\begin{equation}
    A_T(k_*) = A_S(k_*) \, r, \quad A_S(k_*) = 2.0989 \times 10^{-9}.
\end{equation}
For our analysis, we fix $r = 0.035$. In standard single-field slow-roll inflation with Bunch-Davies initial vacuum, the spectral index $n_T$ satisfies the consistency relation $n_T \approx -r/8$ \cite{Liddle:1993fq}, leading to a red-tilted spectrum ($n_T < 0$). However, alternative inflationary models and particle production models allow for deviations, including a blue-tilted spectrum ($n_T > 0$), which have significant theoretical motivation and have been thoroughly analyzed in the literature \cite{Brandenberger:2006xi, Baldi:2005gk, Kobayashi:2010cm, Calcagni:2004as, Calcagni:2013lya, Cook:2011hg, Mukohyama:2014gba, Kuroyanagi:2020sfw}. For our purpose we will keep $n_T$ as a free parameter.

The transfer function $T_T^2(k)$ is determined analytically and numerically as
\begin{equation}
    T_T^2(k) = \Omega_m^2 \left(\frac{g_*(T_{\text{in}})}{g_*^0}\right) 
    \left(\frac{g_{*S}^0}{g_{*S}(T_{\text{in}})}\right)^{4/3} 
    \left(\frac{3j_1(z_k)}{z_k}\right)^2 F(k),
\end{equation}
where $g_*^0 = 3.36$ and $g_{*S}^0 = 3.91$ are the present-day values of the relativistic degrees of freedom. The total matter density parameter is $\Omega_m = 0.31$ \cite{Datta:2022tab}. Here, $j_1(z_k)$ is the spherical Bessel function, with $z_k \equiv k\tau_0$ and $\tau_0 = 2/H_0$. For $z_k \gg 1$, corresponding to the frequencies of interest, the damping factor simplifies to $j_1(z_k) \sim 1/(\sqrt{2}z_k)$. The fitting function $F(k)$ varies depending on the thermal history of the Universe.

In standard cosmology, the fitting function $F(k)$ is given by \cite{Kuroyanagi:2014nba}
\begin{equation}
    F(k)_{\text{standard}} = T_1^2\left(\frac{k}{k_{\text{eq.}}}\right)T_2^2\left(\frac{k}{k_{\text{RH}}}\right),
\end{equation}
whereas in the presence of an intermediate matter domination (IMD) phase, it takes the form
\begin{equation}
\label{eq:FIMD}
    F(k)_{\text{IMD}} = T_1^2\left(\frac{k}{k_{\text{eq.}}}\right)T_2^2\left(\frac{k}{k_{\text{dec}}}\right)T_3^2\left(\frac{k}{k_{\text{dec S}}}\right)T_2^2\left(\frac{k}{k_{\text{RH S}}}\right).
\end{equation}

The key scales appearing above are given by,
\begin{align}
    k_{\text{eq.}} &= 7.1 \times 10^{-2} \, \text{Mpc}^{-1} \cdot \Omega_m h^2, \\
    k_{\text{dec}} &= 1.7 \times 10^{14} \, \text{Mpc}^{-1} \left(\frac{g_{*S}(T_{\text{dec}})}{g_{*S}^0}\right)^{1/6} \left(\frac{T_{\text{dec}}}{10^7 \, \text{GeV}}\right), \\
    k_{\text{RH}} &= 1.7 \times 10^{14} \, \text{Mpc}^{-1} \left(\frac{g_{*S}(T_{\text{RH}})}{g_{*S}^0}\right)^{1/6} \left(\frac{T_{\text{RH}}}{10^7 \, \text{GeV}}\right), \\
    k_{\text{dec S}} &= k_{\text{dec}} \mathcal{D}^{2/3}, \quad k_{\text{RH S}} = k_{\text{RH}} \mathcal{D}^{-1/3}.
\end{align}
Here, $T_{\text{dec}}$ (defined in Eq.~\eqref{eq:Tdec}) and $T_{\text{RH}}$ correspond to the reheating temperatures from HNL and inflaton decay respectively, with $h = 0.7$ being the reduced Hubble parameter. The factor $\mathcal{D}$ accounts for the entropy dilution during the IMD phase, defined in Eq.~\eqref{eq:dilution_factor}.

The fit functions $T_1^2(x)$, $T_2^2(x)$, and $T_3^2(x)$ are defined as
\begin{align}
\label{eq:transfer}
    T_1^2(x) = 1 + 1.57x + 3.42x^2, \quad
    T_2^2(x) = (1 - 0.22x^{3/2} + 0.65x^2)^{-1}, \quad
    T_3^2(x) = 1 + 0.59x + 0.65x^2.
\end{align}

Using Eq.~\eqref{eq:OmegaGW}, the GW spectrum $\Omega_{\text{GW}} h^2$ as a function of the current frequency can be plotted. The spectrum undergoes suppression above a characteristic frequency $f_{\text{sup}}$, given by \cite{Seto:2003kc}
\begin{equation}
\label{eq:fsup}
    f_{\text{sup}} \simeq 
    2.7 \times 10^{-8} \, \text{Hz} 
    \left(\frac{T_{\text{dec}}}{\text{GeV}}\right).
\end{equation}
The suppression factor, $R_{\text{sup}}$, comparing the IMD spectrum to the standard case, reads
\begin{equation}
\label{eq:Rsup}
    R_{\text{sup}} = 
    \frac{\Omega_{\text{GW}}^{\text{IMD}}}{\Omega_{\text{GW}}^{\text{standard}}} 
    \simeq \frac{1}{\mathcal{D}^{4/3}}.
\end{equation}
Ideally when $\mathcal{D}=1$, there is no suppression and $\Omega_{\rm GW}^{\rm IMD}$ approximates $\Omega_{\rm GW}^{\rm standard}$. However, the numerical formulas in Eq.~\eqref{eq:transfer}, especially the $0.59x$ factor in $T_3^2(x)$ produces a bump in the GW spectrum for small values of the dilution factor $\mathcal{D}$ (see Fig. 1 in~\cite{Kuroyanagi:2008ye}). As suggested there-in, we ignore this $0.59x$ factor for $\mathcal{D}\leq 10$, but we still see a small bump when $\mathcal{D}<2$. Therefore, we take a smooth interpolation from $\Omega_{\rm GW}^{\rm IMD}$ to $\Omega_{\rm GW}^{\rm standard}$ as $\mathcal{D}$ varies from 10 to 1,
\begin{equation}
    \Omega_{\rm GW}^{\rm interpolation} = 
    \Omega_{\rm GW}^{\rm IMD} + \left(\Omega_{\rm GW}^{\rm standard} 
    - \Omega_{\rm GW}^{\rm IMD}\right)
    \left[\frac{2}{1 + e^{-p(\mathcal{D})}} - 1\right],
\end{equation}
where $p(\mathcal{D})$ depends on the dilution. For $p=0$, the spectrum is described by $\Omega_{\rm GW}^{\rm IMD}$ while for large $p$ values, the spectrum smoothly transitions to $\Omega_{\rm GW}^{\rm standard}$. We take the form of the parameter $p(\mathcal{D})=10\, e^{-(\mathcal{D}-1)}$ for our analysis.

\begin{figure}[t!]
    \centering
    \includegraphics[width=0.49\linewidth]{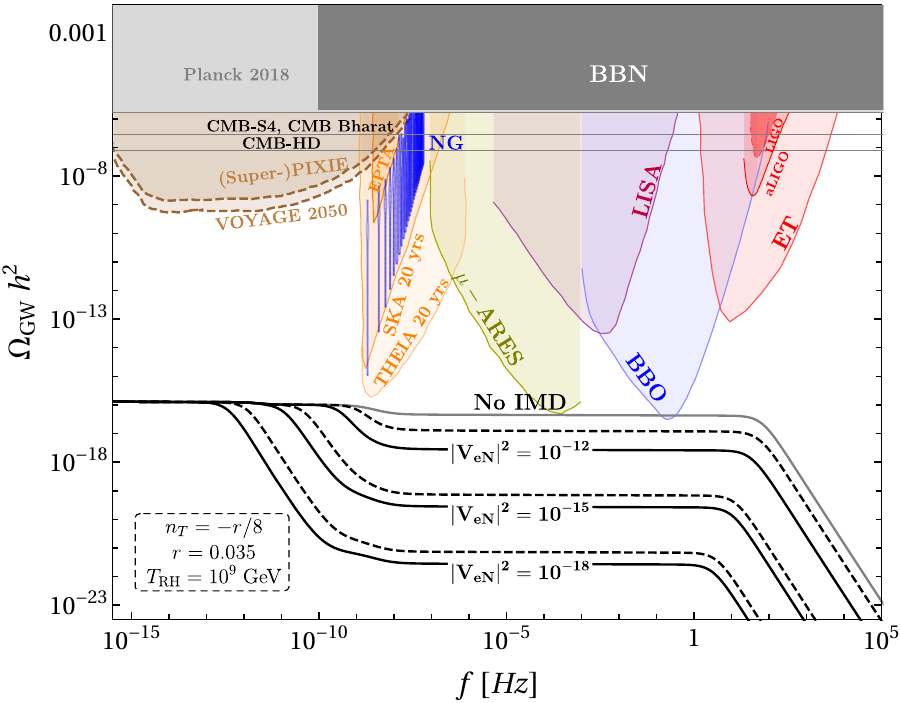}
     \includegraphics[width=0.49\linewidth]{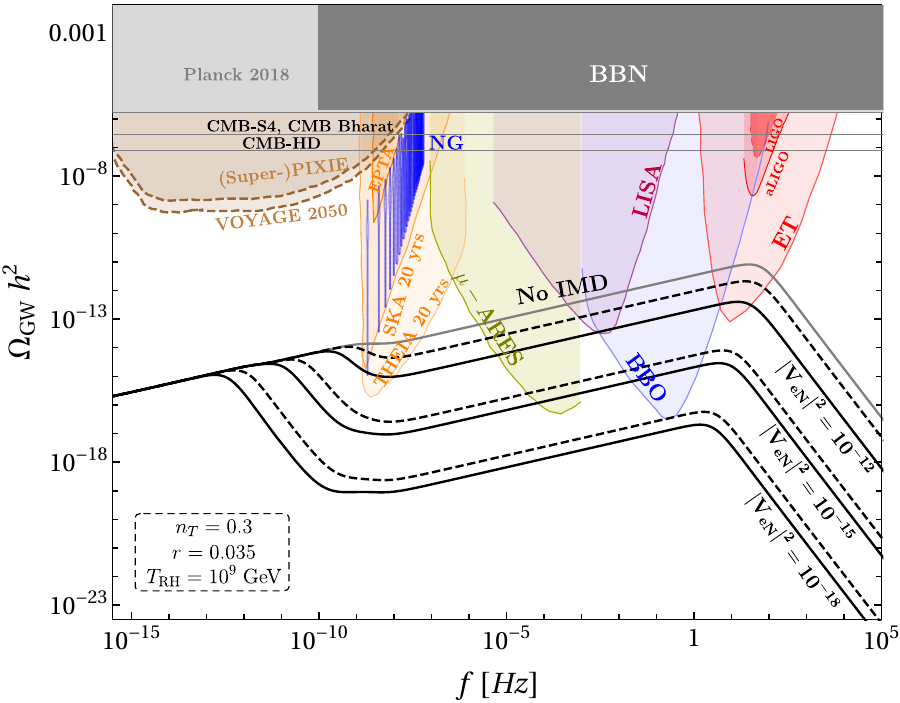}
    \caption{Example GW spectra demonstrating the effect of varying the HNL mass $m_N$ and the active-sterile mixing $|V_{eN}|^2$ on suppression depicted by the black solid ($m_N = 3$~GeV) and dashed ($m_N = 5$~GeV) curves. Left and right plots correspond to spectral indices $n_T = -r/8$ (implying slow-roll inflation) and $0.3$ (blue-tilted), respectively. The gray solid curve indicates the case without HNLs and thus no intermediate matter domination. Other parameters we have taken are $r=0.035, T_{\rm RH}=10^9$ GeV. The dark/light gray regions on top are the current $\Delta N_{\rm eff}$ bounds from BBN and CMB respectively, that rule out $\Omega_{\rm GW}h^2\gtrsim10^{-6}$. The red region corresponding to LIGO is also ruled out. The other colored regions represent expected sensitivities of future GW observations.}
    \label{fig:specn05Th}
\end{figure}

In Fig.~\ref{fig:specn05Th}, we show example GW spectra for various values of $m_N$ and $|V_{e N}|^2$. The left plot shows the spectra for $n_T = -r/8$ corresponding to slow-roll inflation, with $r = 0.035$. The right plot shows blue-tilted spectra with $n_T = 0.3$. The standard PGW spectrum is represented by the solid gray curve, corresponding to no intermediate matter domination. The dashed and solid black curves are the suppressed spectra for $m_N=5$~GeV and $3$~GeV respectively. We take $T_{\rm RH} = 10^9$~GeV for both the plots and demonstrate the effects of $n_T$, $m_N$ and $|V_{eN}|^2$ on the suppression of the spectra. 

The shaded colored regions represent the noise curves for various existing and upcoming gravitational wave experiments, including Laser Interferometer Gravitational-Wave Observatory (LIGO)~\cite{LIGOScientific:2016aoc, LIGOScientific:2016sjg}, Advanced LIGO (aLIGO)~\cite{LIGOScientific:2014pky}, Big Bang Observer (BBO)~\cite{Yagi:2011wg}, Laser Interferometer Space Antenna (LISA)~\cite{2017arXiv170200786A}, Einstein Telescope (ET)~\cite{Punturo_2010}, THEIA~\cite{Garcia-Bellido:2021zgu}, $\mu$ARES~\cite{Sesana:2019vho}, European Pulsar Timing Array (EPTA)~\cite{Antoniadis:2023ott}, and Square Kilometer Array (SKA)~\cite{Weltman:2018zrl}. The blue violin curves show recent NANOGrav (NG) results~\cite{NANOGrav:2023gor}. Future missions such as Super-PIXIE~\cite{Kogut:2024vbi} and VOYAGE 2050~\cite{Basu:2019rzm} are designed to detect gravitational wave signals through spectral distortions in the cosmic microwave background.
The current bounds on $\Delta N_{\rm eff}$ from BBN and CMB are shown as gray boxes at the top while the future projections are shown as horizontal gray lines. 

\begin{figure}[t!]
    \centering
    \includegraphics[width=0.52\linewidth]{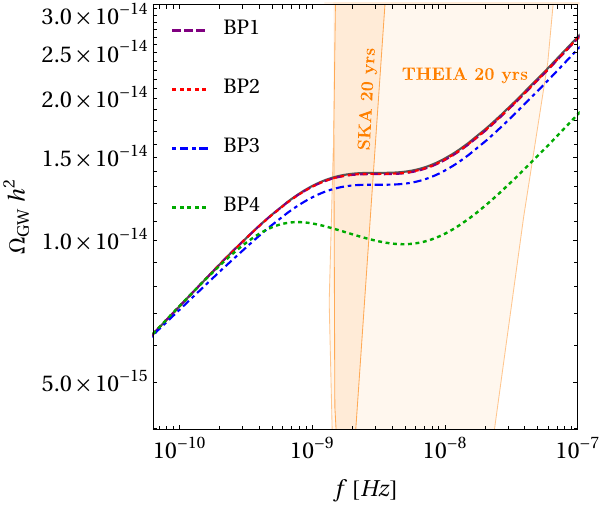}
    \includegraphics[width=0.465\linewidth]{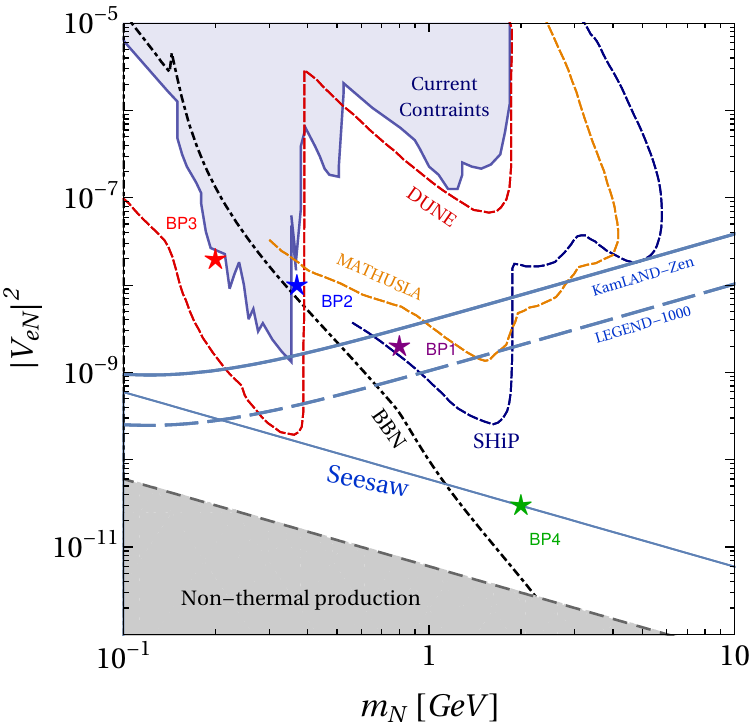}
    \caption{Left: GW spectra for the benchmark points in Table~\ref{tab:BPvalues}, with the primordial tensor spectral index $n_T = 0.3$ and the tensor-to-scalar ratio $r = 0.035$. The colored regions represent the the sensitivity of the future GW observations at SKA and THEIA. Right: Current constraints (blue region) and future sensitivities (colored curves labeled DUNE, MATHUSLA, SHiP) of direct HNL searches on the active-sterile mixing $|V_{eN}|^2$ as a function of the HNL mass $m_N$. Also shown are the current constraint from $0\nu\beta\beta$ decay at KamLAND-Zen and the future $0\nu\beta\beta$ decay sensitivity at LEGEND-1000. The region to the left of the curve labeled 'BBN' is disfavored by Big Bang Nucleosynthesis and HNLs are not thermally produced in the gray region. The line labeled 'Seesaw' indicates successful neutrino mass generation, $m_\nu = |V_{eN}|^2 m_N = 0.1$~eV, in the canonical seesaw. The stars indicate the same benchmark points from Table~\ref{tab:BPvalues} as in the left panel.}
    \label{fig:specBP}
\end{figure}

In Fig.~\ref{fig:specBP}, the left panel shows the GW spectra for the benchmark points given in Table~\ref{tab:BPvalues}, while the right panel shows the benchmark points in the HNL parameter space. The purple dashed, blue dot-dashed, red dotted and green-dotted lines show the damped GW spectra with HNL domination for the benchmark points BP1, BP2, BP3 and BP4. The shaded region indicates the GW sensitivity curves of the SKA and THEIA experiments. In the right panel, colored dashed lines represent the projected sensitivities of DUNE, MATHUSLA, and SHiP, while the dot-dashed black line shows the BBN bound on the lifetime of HNL~\cite{Boyarsky:2020dzc}. Below the BBN line, the lifetime of HNL exceeds $\sim$ 0.02 s and for $m_N$ larger than the mass of pion, the decay of HNL to pions during BBN can lead to  over-production of primordial helium-4. Hence this region must be excluded. The gray region at the bottom represents non-thermal production of HNL where Eq.~\eqref{eq:nonThCond} is not satisfied and we have not studied that region in this paper. BP1 lies within the sensitivity of SHiP and LEGEND-1000, while BP2 and BP3 fall within DUNE's, indicating potential signals in both GW and laboratory experiments. Note that BP3 is excluded by BBN, however this is a weak bound due to inherent uncertainties in its calculation. Recent studies~\cite{Yeh:2024ors, Chen:2024cla} provide more stringent constraints on the parameter space $m_N$ vs. $|V_{eN}|^2$ by analyzing the effects of matter-domination during BBN in light of CMB and BBN data. Future experiments such as CMB-S4 are expected to provide significantly improved measurements, potentially tightening these constraints further.
\begin{table}[t!]
\centering
\begin{tabular}{@{}|c|c|c||c|c|c|c|c|@{}}
\hline
\multirow{3}{*}{Benchmark} & \multirow{3}{*}{$\,\,m_N\;(\text{GeV})\,\,$} & \multirow{3}{*}{$|V_{eN}|^2$} & \multicolumn{5}{c|}{SNR$_{\rm rel}\times 100\%$} \\
\cline{4-8}
 & & & \multicolumn{2}{c|}{$n_T = -r/8$} & \multicolumn{3}{c|}{$n_T = 0.3$} \\
\cline{4-8}
 & & & BBO & $\mu-$ARES & THEIA & LISA & ET \\
\hline
BP1 & $0.80$ & $2\times 10^{-9}$  & 0.8\%  & 0.8\%  & 0.5\%  & 0.8\%  & 0.8\%  \\
BP2 & $0.37$ & $1\times 10^{-8}$  & 0.5\%  & 0.5\%  & 0.4\%  & 0.5\%  & 0.5\%  \\
BP3 & $0.20$ & $2\times 10^{-8}$  & 5.8\%  & 5.8\%  & 5.5\%  & 5.8\%  & 5.8\%  \\ 
BP4 & $2.00$ & $3\times 10^{-11}$ & 31.7\% & 31.7\% & 22.4\% & 31.7\% & 31.7\% \\ 
\hline
\end{tabular}
\caption{Benchmark points and their SNR$_{\rm rel}$ projections in THEIA, $\mu-$ARES, LISA, BBO, and ET. BP1 comes within sensitivity of SHiP and LEGEND-1000, while BP2 and BP3 are sensitive in DUNE. The benchmark BP4 satisfies the type-I seesaw relation and generates light neutrino masses. The value of tensor-to-scalar ratio is $r = 0.035$ and inflationary reheating temperature is $T_{\rm RH} = 10^9$ GeV.}
\label{tab:BPvalues}
\end{table}

\subsection{Signal-to-noise ratio}

Interferometers serve as precision instruments to measure displacements caused by gravitational waves (GWs), expressed in terms of the dimensionless strain-noise $h_{\text{GW}}(f)$. The strain-noise is intrinsically connected to the amplitude of gravitational waves and can be translated into the corresponding energy density spectrum. This relationship is given by
\begin{equation}
    \Omega_{\text{exp}}(f)h^2 = \frac{2\pi^2 f^2}{3H_0^2} h^2_\text{GW}(f) h^2,
\end{equation}
where $H_0$ is the present-day Hubble expansion rate, defined as $H_0 = h \times 100 \, \frac{\text{km/s}}{\text{Mpc}}$, with $h$ denoting the dimensionless Hubble parameter.

\begin{figure}[t!]
    \centering     
    \includegraphics[width=0.6\linewidth]{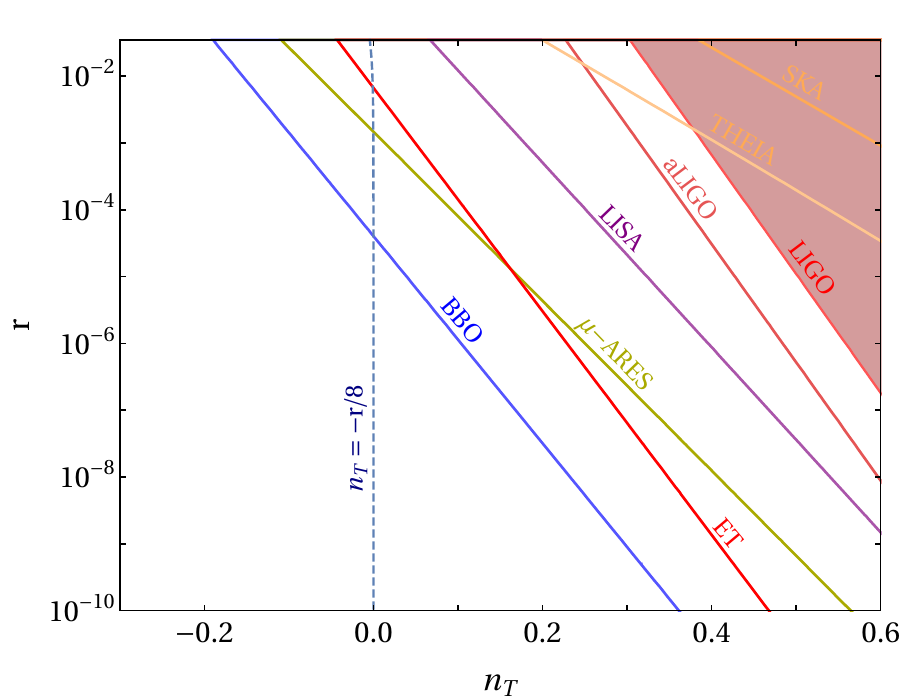} 
    \caption{Future detectability of standard PGW spectra (no intermediate matter domination) across GW experiments. Regions to the right of each colored line indicate where SNR $>10$ in the $(n_T, r)$ parameter space for different GW experiments. The dashed line $n_T \approx -r/8$ denotes standard slow-roll inflation. We fix $T_{\text{RH}} = 10^9$ GeV.}
    \label{fig:nTr}
\end{figure}

To determine the detectability of the primordial gravitational wave background, we compute the signal-to-noise ratio (SNR) based on the experimental sensitivity curve $\Omega_{\text{exp}}(f)h^2$, which may represent current measurements or future projections. The SNR is given by
\begin{equation}
\label{eq:SNR}
    \text{SNR} \equiv \sqrt{\tau \int_{f_{\text{min}}}^{f_{\text{max}}} df \left(\frac{\Omega_\text{GW}(f)h^2}{\Omega_{\text{exp}}(f)h^2}\right)^2},
\end{equation}
where $\tau$ is the total observation time and the integral runs over the frequency range $[f_{\text{min}}, f_{\text{max}}]$. For the purposes of this analysis, we assume an observation time of $\tau = 4$ years and set $h = 0.7$. A detection threshold is established by requiring a signal-to-noise ratio of $\text{SNR} \geq 10$ \cite{Caprini:2018mtu, Breitbach:2018ddu}. 
In Fig.~\ref{fig:nTr}, we show regions with SNR $>10$ in the parameter space spanned by $n_T$ and $r$ for different future GW experiments. Here, we only consider the standard primordial GW spectra and assume a reheating temperature of $T_\text{RH} = 10^9$~GeV. The vertical dashed line corresponding to $n_T \approx -r/8$ represents the standard slow-roll inflation, which can be tested only with BBO, ET, and $\mu-$ARES. We see that $n_T > 0$ is required to get a detectable GW signal in LISA, THEIA, and SKA. The figure illustrates that primordial GWs are in principle detectable over a wide range of parameters. While positive values of $n_T$ are preferred, this includes predictions from standard slow-roll inflation. Such a signal is then further modified by the IMD driven by a right-handed neutrino species in the early universe as our main effect of interest.

\subsection{Bounds from dark radiation}
\label{sec:delNeff}

\begin{figure}[t!]
    \centering     
    \includegraphics[width=0.38\linewidth]{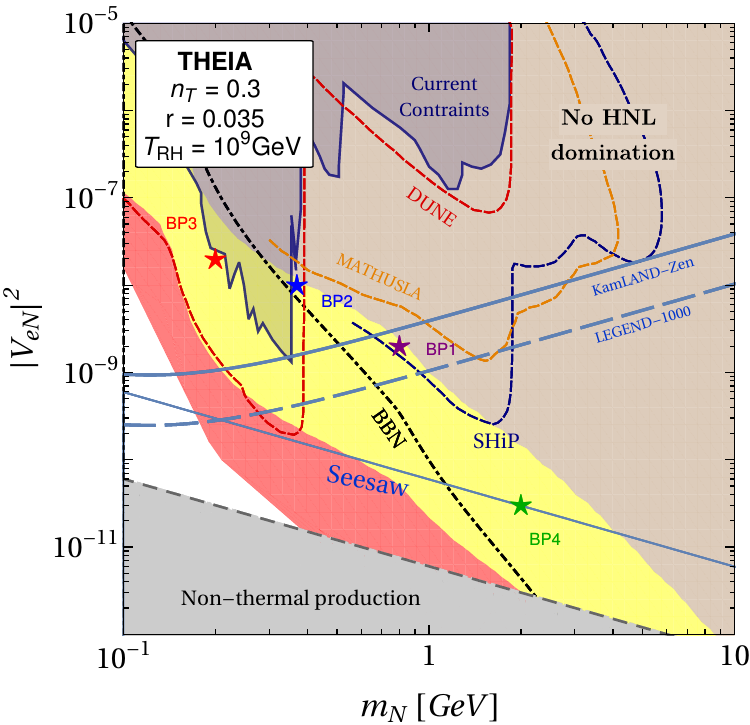}  
    \includegraphics[width=0.38\linewidth]{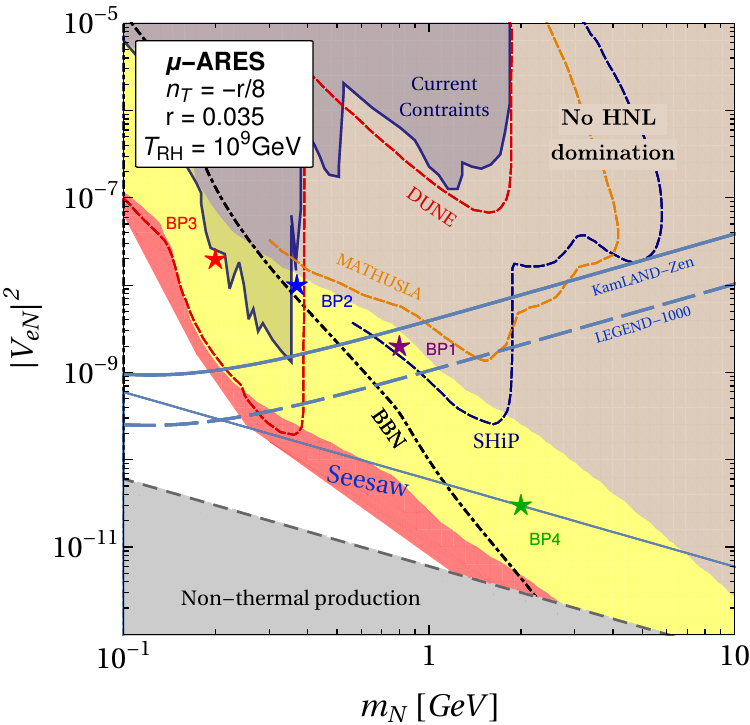}
    \includegraphics[width=0.38\linewidth]{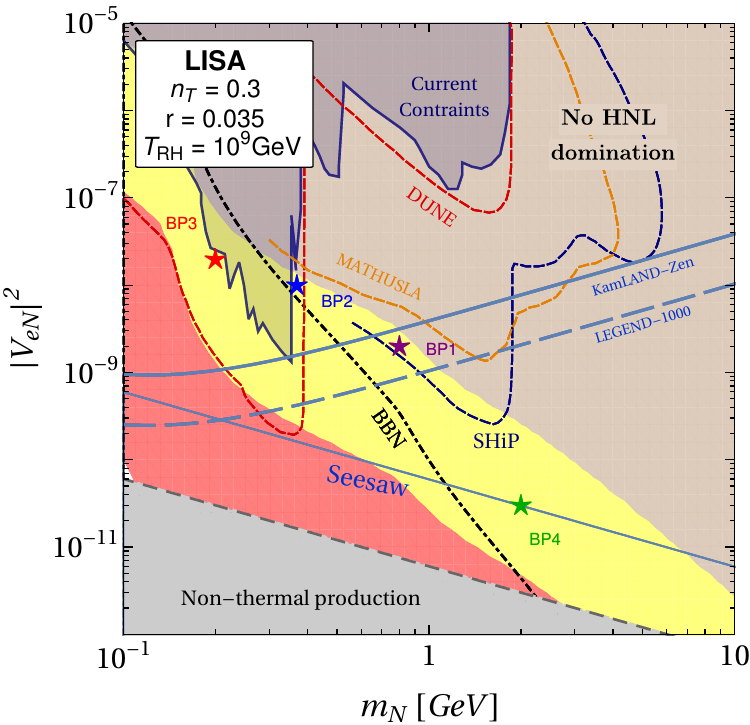}
    \includegraphics[width=0.38\linewidth]{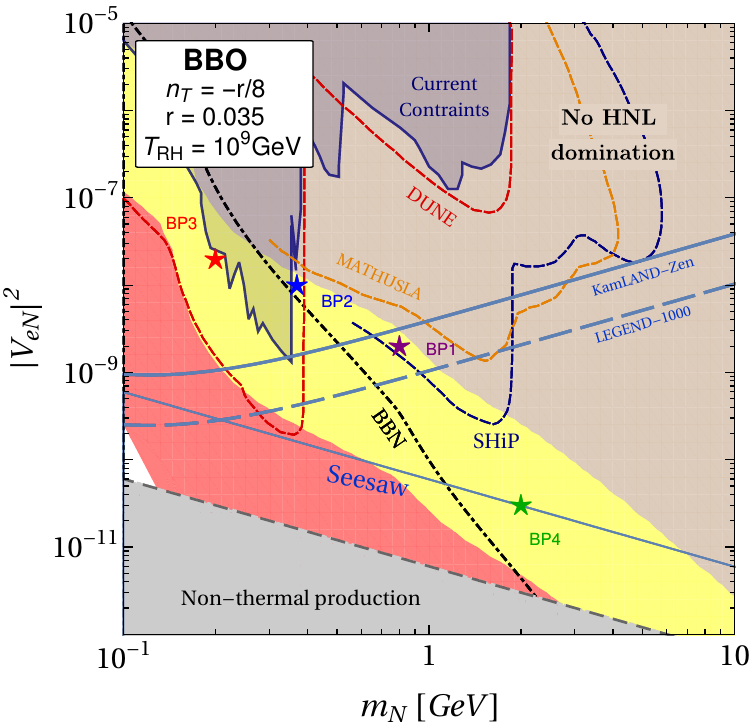}
    \includegraphics[width=0.38\linewidth]{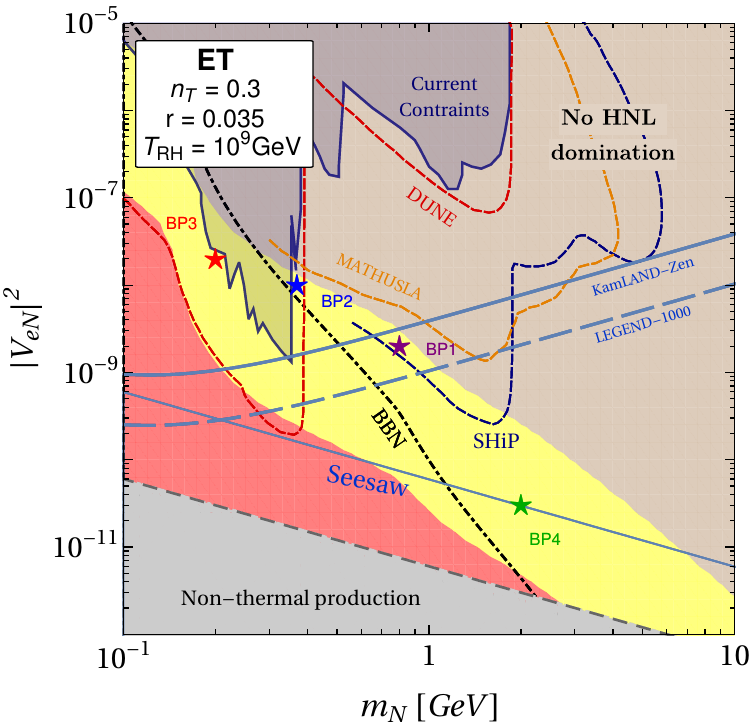}
    \caption{Projected sensitivities for HNLs with mass $m_N$ and active-sterile mixing $|V_{eN}|^2$ at different future GW observations, in terms of the relative signal-to-noise ratio SNR$_\text{rel}$: SNR$_\text{rel} < 0.003$ (no HNL domination, brown region); $0.003\leq \text{SNR}_\text{rel}\leq 0.9$ (weak domination, yellow region); SNR$_\text{rel} \geq 0.9$ (strong domination, red region). A white region in the bottom left indicates an absolute SNR$_\text{IMD} < 10$ that is undetectable in the given GW observation. For comparison, each panel shows the current constraints and future sensitivities from direct and $0\nu\beta\beta$ decay searches for HNLs, as described in Fig.~\ref{fig:specBP}.}
    \label{fig:SNR1}
\end{figure}
We present here the effective constraints on dark radiation arising due to BBN and CMB decoupling. Since gravitons behave as massless radiation-like degrees of freedom, the primordial gravitational wave (GW) energy density considered should remain below the observational limits on any such additional dark radiation. This is conventionally expressed in terms of deviations in the effective number of relativistic degrees of freedom, $\Delta N_\text{eff}$. The constraint on $\Omega_\text{GW}(f)$ at the time of recombination is usually expressed in terms of~\cite{Maggiore:1999vm}
\begin{equation}
\label{eq:darkrad}
    \int_{f_\text{min}}^{\infty} \frac{\mathrm{d}f}{f} \, \Omega_\text{GW}(f) h^2 \leq 5.6 \times 10^{-6} \, \Delta N_\text{eff}. 
\end{equation}
Here, the lower bound of integration, $f_\text{min}$, typically differs depending on the cosmological epoch under consideration. Particularly, we know that for BBN, $f_\text{min} \sim 10^{-10} \, \mathrm{Hz}$, while for CMB constraints, it is approximately taken to be $f_\text{min} \sim 10^{-18} \, \mathrm{Hz}$. Nonetheless, as for all practical purposes, such as when comparing multiple GW spectra or analyzing their peak contributions, it is more than sufficient to approximate these constraints by ignoring the detailed frequency dependence and instead focusing only on the total energy density at the spectral peak. This simplifies a lot and consequently leads to the expression
\begin{equation}
\label{eq:darkrad2}
    \Omega_\text{GW}^\text{Peak} h^2 \leq 
    5.6 \times 10^{-6} \, \Delta N_\text{eff}. 
\end{equation}
For our present analysis, we incorporate current bounds on $\Delta N_\text{eff}$ derived from BBN observations and the PLANCK 2018 data from references~\cite{Planck:2018vyg}. Additionally, we explore the sensitivity reach of upcoming CMB experiments, including CMB-S4~\cite{CMB-S4:2020lpa, CMB-S4:2022ght}, CMB-Bharat~\cite{CMB-bharat}, and CMB-HD~\cite{Sehgal:2019ewc, CMB-HD:2022bsz}. These we expect to significantly refine the constraints on $\Delta N_\text{eff}$ or discover extra dark radiation and, consequently, provide sensitivity reaches on the primordial GW energy density.
\begin{figure}[t!]
    \centering     
    \includegraphics[width=0.38\linewidth]{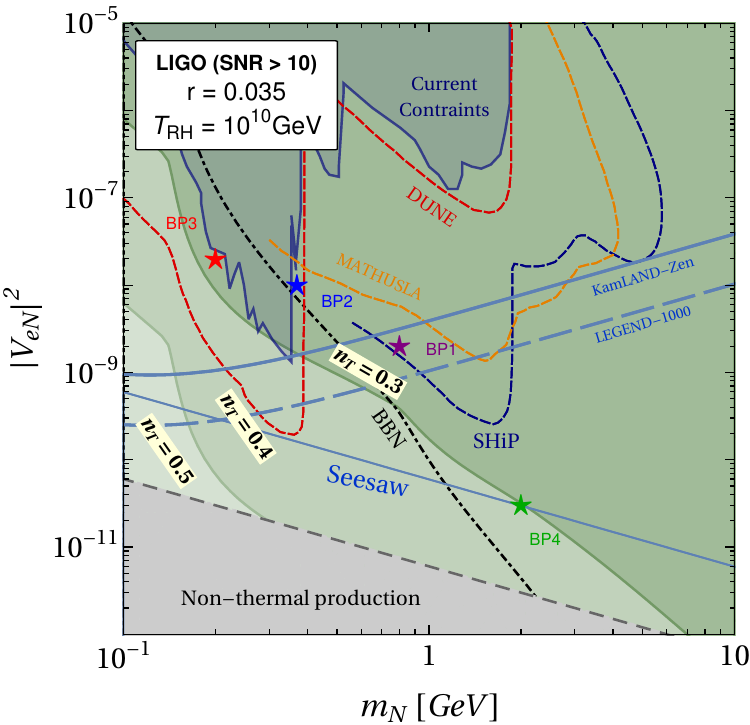}
    \includegraphics[width=0.38\linewidth]{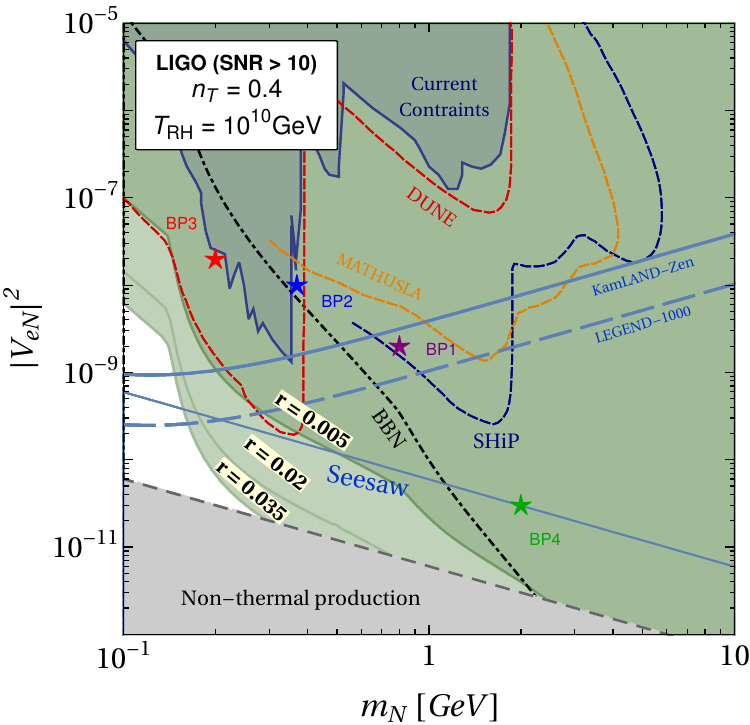}
    \includegraphics[width=0.38\linewidth]{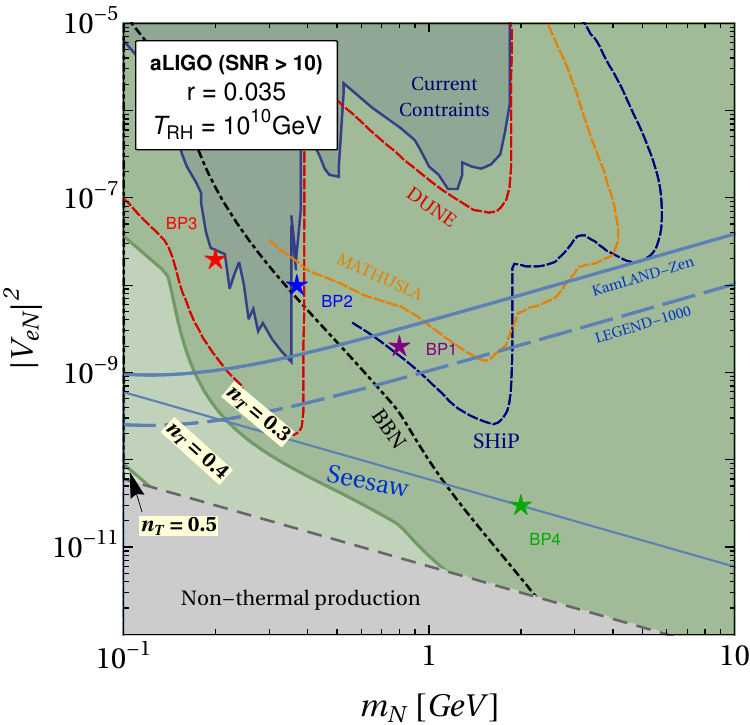}
    \includegraphics[width=0.38\linewidth]{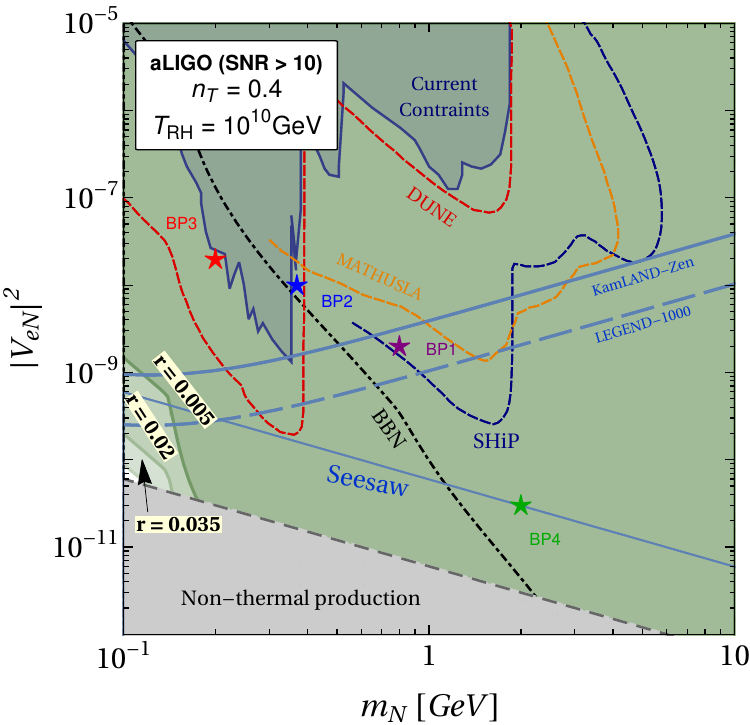}
    \caption{Regions with SNR $ > 10$ (green) for LIGO (top) and advanced LIGO (bottom) in the RHN parameter space. The left panels show the effect of varying the spectral index, $n_T = 0.3$, $0.4$ and $0.5$, keeping the tensor-to-scalar ratio $r = 0.035$. The right panels show the effect of varying $r = 0.035$, $0.02$ and $0.005$, for a fixed value of $n_T = 0.4$. All plots assume a constant reheating temperature $T_\text{RH} = 10^{10}$~GeV.}
    \label{fig:LigoBound}
\end{figure}
\subsection{Results}
\label{sec:results}

We present the projected sensitivities of future GW experiments in the parameter space of the HNL. In Fig.~\ref{fig:SNR1}, we show laboratory searches for HNL described in Sec.~\ref{sec:HNL searches}, along with contours of the relative SNR defined as
\begin{equation}
\label{eq:delSNRoverSNR}
    \text{SNR}_{\rm rel} \equiv 
    \frac{\text{SNR}_{\rm std.}-\text{SNR}_{\rm IMD}}{\text{SNR}_{\rm std.}}
\end{equation}
for various GW experiments where SNR$_{\rm std.}$ is the SNR of the standard spectra, which is a constant for a given GW experiment given fixed values of $r,n_T$ and $T_{\rm RH}$. SNR$_{\rm IMD}$ is the SNR of the suppressed GW spectra due to HNL domination. The quantity SNR$_{\rm rel}$ signifies the deviation of the GW spectra from the expected standard spectra such that larger value of SNR$_{\rm rel}$ correspond to larger suppression. It varies between $0$ and $1$ with $0$ signifying no suppression and $1$ signifying huge suppression or no detection prospect. Another benefit of showing SNR$_{\rm rel}$ instead of just showing SNR$_{\rm IMD}$ is that the effects of $r$, $n_T$ and $T_{\rm RH}$ are canceled out in SNR$_{\rm rel}$. For $\mu-$ARES and BBO, we take $n_T= -r/8$ corresponding to slow-roll inflation, with $r = 0.035$. However, for THEIA, LISA and ET, we take $n_T=0.3$. We find that for $n_T=0.3$, all the benchmark points are ruled out by LIGO data if $T_{\rm RH}\gtrsim10^9$ GeV. We take $T_{\rm RH}=10^9$ GeV for all the plots in Fig.~\ref{fig:SNR1}. The light brown region signifies the scenario where HNLs do not dominate in the early Universe, hence SNR$_{\rm rel}$ $\sim0$ (due to numerical uncertainties, we ignore SNR$_{\rm rel}$ $<0.003$), i.e. we can not differentiate the GW spectra from the standard spectra. Yellow region is where $0.003\leq\Delta$SNR $\leq0.9$ while red region is where SNR$_{\rm rel}$ $>0.9$. We put a threshold, SNR $> 10$ for these plots signifying a 5$\sigma$ detection prospect at all these GW experiments. The white region in the bottom left of each plot represents SNR$_{\rm IMD}<10$, i.e. GW is difficult to detect in this region due to large suppression. 

The value of $n_T$ taken for each plot is shown in the plot. The four benchmark points BP1, BP2, BP3 and BP4 are testable in both GW and laboratory experiments. The corresponding values of SNR$_{\rm rel}$ for the benchmark points at various GW experiments are shown in Table~\ref{tab:BPvalues}. We find that SNR$_{\rm rel}<1\%$ in case of BP1 and BP2 in all GW experiments, implying marginal complementarity of GW experiments with SHiP and DUNE. However we see there is a region in the parameter space ($0.6\text{ GeV}\lesssim m_N\lesssim1\text{ GeV}$ and $|V_{eN}|\sim10^{-9}$) with large SNR$_{\rm rel}$, which is also sensitive to LEGEND-1000. GW searches can also probe regions of the parameter space where no laboratory searches reach. For example GW can probe the seesaw line (solid blue line) where active neutrino masses are successfully generated via the seesaw mechanism, as depicted by the benchmark BP4. Note that for all the benchmark points, SNR$_{\rm rel}$ in THEIA is smaller than that in other experiments. This is due to the fact that for all the benchmark points, the frequency of suppression $f_{\rm sup}$ defined in Eq.~\eqref{eq:fsup} lie within the frequency range of THEIA.
\subsection{Constraints from LIGO}
The Laser Interferometer Gravitational wave Observatory (LIGO) is a large-scale experiment designed to detect high-frequency gravitational waves in the range of approximately $30$–$207$ Hz. Its upgraded version, Advanced LIGO (aLIGO), features significantly enhanced sensitivity over a broader frequency band, extending from $\sim 20$ to $450$~Hz. Data from aLIGO's fourth observational run (O4) is expected to be released by the end of 2025. Figure~\ref{fig:LigoBound} shows the current and projected constraints from LIGO and aLIGO in the parameter space of $m_N$ and $|V_{eN}|^2$. The green shaded regions in the upper panels show SNR $ > 10$ constraints for LIGO, while the lower panels display projected regions for aLIGO. The left column explores variations in the tensor spectral index $n_T$ for fixed $r$, whereas the right column varies $r$ at fixed $n_T$. In all plots, the reheating temperature $T_\text{RH} = 10^{10}$~GeV is assumed. It can be clearly seen that higher value of $n_T$ and $r$ increases the amplitude of GW hence increasing SNR in the LIGO and aLIGO. We find that for $n_T = 0.3$ and $T_\text{RH} \leq 10^9$~GeV, LIGO does not put any constraint on the considered parameter space. Hence, the HNL parameter space is allowed in Fig.~\ref{fig:SNR1} for $T_{\rm RH} = 10^9$~GeV.


\section{Discussion and Conclusions}
\label{sec:conclusion}

Inflationary gravitational waves offer us a unique window into heavy and high energy particle physics that is otherwise inaccessible to terrestrial experiments~\cite{Dasgupta:2022isg, Bhaumik:2022pil, Barman:2022yos, Ghoshal:2022jdt, Dunsky:2021tih, Bernal:2020ywq, Ghoshal:2020vud}. In this work, we show that GWs can probe long-lived RHNs by discerning deviations from the standard thermal history due to a brief period of early matter domination during the pre-BBN era. Starting from a thermal initial abundance, the suppressed decay rate of RHNs enables them to survive and dominate the energy budget of the universe temporarily, leaving imprints in the inflationary GW spectrum in the form of a characteristic dip the amplitude. The entire setup is therefore described by a minimal set of independent parameters: the mass of RHNs, $m_N$, and the active-sterile mixing, $|V_{eN}|^2$. We find a novel complementary signature for long-lived RHNs in GW searches THEIA, LISA, ET, $\mu-$ARES and BBO, by studying the effect on the GW spectral shape caused by a period of early matter domination due to long-lived RHN, see Fig.~\ref{fig:specn05Th}. We find that ET, THEIA and LISA will not be able to detect significant suppression of PGW spectra for $n_T\lesssim -r/8$ with $r = 0.035$. Specifically, THEIA and LISA require $n_T\gtrsim0.3$ for detection. However, in this case, LIGO rules out most of the parameter space considered, if $T_{\rm RH}\gtrsim10^9$ GeV. In Fig.~\ref{fig:nTr} we show the regions with SNR $>10$ for standard GW spectra without IMD in the parameter space ($n_T, r$). We find that very small parameter space gives detectable signal for experiments like BBO, ET and $\mu$-ARES under the standard slow roll scenario.  For other GW experiments such as LISA, THEIA, and ET, a blue-tilted PGW spectrum ($n_T > 0$) will be detectable.

We compare the resulting sensitivity with that of laboratory searches for RHNs, namely beam dump experiments and $0\nu\beta\beta$ decay, where long-lived RHNs can be searched for experimentally in light-dark sector searches involving intensity, lifetime and beam dump experiments. We explored the parameter space in Fig.~\ref{fig:SNR1}, highlighting the projected sensitivity of long-lived particle searches at DUNE, MATHUSLA, and SHiP as well as the $0\nu\beta\beta$ decay search at LEGEND-1000. Interestingly, we demonstrated overlapping regions of sensitivity in such searches with detectable signals within four years of exposure for the GW detectors considered. These may serve as complementary signatures for RHNs and give a deep insight into the mechanism of neutrino mass generation. This is expected to enhance the existing complementarity between laboratory searches \cite{Bolton:2022tds} and would provide direct evidence of RHNs in early cosmology. The RHN mass and mixing range probed by laboratory experiments and GW detectors is $0.2~\text{GeV} \lesssim m_N \lesssim 2~\text{GeV}$ and $10^{-10} \lesssim |V_{eN}|^2 \lesssim 10^{-7}$, as evident from Fig.~\ref{fig:specBP}.

In addition, GW detectors can probe regions of parameter space not reachable by laboratory experiments, particularly the canonical seesaw regime of light neutrino mass generation. This regime is difficult to probe with direct laboratory and GW signals may not only be able to shed light on this important seesaw regime but also various well-motivated dark sectors, especially in the weakly coupled regimes, such as the neutrino-portal portal freeze-in dark matter \cite{Barman:2022scg}, axion-portal RHNs \cite{Deppisch:2024izn} and low scale leptogenesis \cite{Klaric:2020phc}. We plan to look into such analyses in a future publication. We emphasize again that hunting for apparently disjointed signals in laboratory experiments and cosmological probes will help us to break degeneracies beyond the SM theories of cosmology that involve multiple energy scales and heavy scales, thus providing a possible pathway to the community and a promising way to search for new physics in the upcoming years.

\section*{Acknowlegement}
Z.~A.~B. is supported by the IoE fund at IIT Bombay. L.~M. is supported by an UGC scholarship. F.~F.~D. acknowledges support from the UK Science and Technology Facilities Council (STFC) via the Consolidated Grant ST/X000613. AG thanks Stefan Pokorski and Deep Ghosh for discussions.

\bibliography{bibliography}
\bibliographystyle{JHEP}
\end{document}